\documentclass[a4paper,twocolumn,fleqn]{article} 

\usepackage{ist}
\usepackage{ctable} %

\usepackage{color}

\newcommand{\comment}[1]{}
\title{Extending the Unmixing methods to Multispectral Images}

\author{Jizhen Cai,${}^1{}^2{}^3{}$ Hermine Chatoux,${}^1$ Clotilde Boust,${}^2{}^3{}$ Alamin Mansouri${}^1$  \\  \\
${}^1$        Laboratoire Imagerie et Vision Artificielle, Universite Bourgogne Franche-Comt\'e, France   \\
${}^2$        Le Centre de Recherche et de Restauration des Mus\'ees de France, France\\
${}^3$        Centre National de la Recherche Scientifique, France
}

\date{} 

\begin{document} 

\maketitle 

\thispagestyle{empty} 


\begin{abstract}
In the past few decades, there has been intensive research concerning the Unmixing of hyperspectral images. Some methods such as NMF, VCA, and N-FINDR have become standards since they show robustness in dealing with the unmixing of hyperspectral images. However, the research concerning the unmixing of multispectral images is relatively scarce. Thus, we extend some unmixing methods to the multispectral images. In this paper, we have created two simulated multispectral datasets from two hyperspectral datasets whose ground truths are given. Then we apply the unmixing methods (VCA, NMF, N-FINDR) to these two datasets. By comparing and analyzing the results, we have been able to demonstrate some interesting result for the utilization of VCA, NMF, and N-FINDR with multispectral datasets. Besides, this also demonstrates the possibilities in extending these unmixing methods to the field of multispectral imaging.
\end{abstract}

\section{Introduction}
\label{sec:intro}

Hyperspectral imaging camera captures the information across the electromagnetic spectrum including ultraviolet, visible, and infrared \cite{ferreira2017evaluation, yokoya2011coupled}. The broad spectral range of hyperspectral imaging gives it an incomparable advantage in detection and classification in domains such as food quality control \cite{gowen2007hyperspectral}, coal mining \cite{krupnik2019close}, and ecosystem studies \cite{kokaly2009characterizing}. 

Multispectral imaging has some similarities with hyperspectral imaging, yet they have many differences in essence and applications. One difference is the volume of the data. Hyperspectral imaging usually has around 200 spectral bands, while there usually exists only a few important bands (usually between 4 and 20) in multispectral imaging. This feature makes it much faster to obtain multispectral data than hyperspectral. In fact, as Qin \textit{et al.} have pointed out, in real cases, hyperspectral images are usually used as fundamental datasets from which to determine optimal wavebands that can be used by a multispectral imaging solution for a particular application \cite{qin2013hyperspectral}. In that case, the band width of multispectral images will be larger than the corresponding hyperspectral images because multispectral images have fewer bands in the same range of wavelengths.

An important use of hyperspectral processing is unmixing. It is defined as the process of separating the spectral signatures (endmembers) and respective proportions of each endmember (abundances) from hyperspectral images \cite{bioucas2012hyperspectral}. Many Unmixing methods such as Vertex Component Analysis (VCA), N-FINDR, and Nonnegative Matrix Factorization (NMF) have been validated for hyperspectral images \cite{nascimento2005vertex, winter2004proof, xiong2011fast, rajabi2014spectral}.

However, in real situations, due to various constraints, we might only have multispectral images. In fact, multispectral images are more commonly used, because hyperspectral devices are more expensive and require more processing. As Hruska \textit{et al.} have mentioned, hyperspectral imaging can require more scheduling of time in advance to use advanced, high quality systems such as AVIRIS and HyMap. Moreover, the cost for hyperspectral images is also much higher (around 10 times higher) than that for multispectral images \cite{hruska2012radiometric}.

Nowadays, the research concerning the unmixing of multispectral imaging is becoming increasingly pertinent. On the one hand, there are more and more people using multispectral imaging, because it is more cost-effective and time-efficient than hyperspectral imaging. On the other hand, recent research still concentrates on hyperspectral imaging like the classification of hyperspectral imaging with data mining methods including KNN \cite{huang2016spectral}, K-MEANS \cite{haut2017cloud}, and SVM \cite{mercier2003support}. There are very few research focusing on Multispectral Unmixing.


\section{Unmixing methods}
Although there are several different hyperspectral unmixing methods, in this article we focus on three Unmixing methods: VCA, N-FINDR, NMF. This choice is based on two considerations. Firstly, these three unmixing methods are the most widely used for unmixing. The performance of these three methods on the hyperspectral images prove to be robust and reliable. Secondly, these three methods have differences in their basic assumptions and process. For example, VCA finds the most suitable endmembers by iteratively  projecting data onto the direction orthogonal to the subspace which is already spanned by the endmembers \cite{nascimento2005vertex}. Meanwhile, the idea of N-FINDR is to find the endmembers that will form the largest volume in the N-dimension \cite{winter1999n}. These differences give each method unique advantages depending on the real situations.

VCA assumes the existence of pure pixels in the dataset. Based on that, the endmembers found will always be situated at the vertices of a simplex. VCA has to generate random vectors when starting the extraction of endmembers. According to Chang, such randomness in the initialization made the results of VCA less repeatable at times \cite{chang2013hyperspectral}. 

For N-FINDR, this algorithm is based on the idea that in the N spectral dimensions, the number of N-Volume is gigantic. However, the one that is formed by the purest endmembers will be the largest. Starting with random sets of pixels, it gradually iterates until the largest volume is found. N-FINDR shares a similar feature with VCA in randomness in initial values, which may lead to problems in repeatability. 

As for NMF, its algorithm only allows non-negative values in the procedure of Unmixing. The feature of non-negative values in the matrices makes the result an additive combination of all parts of the images. Such a feature permits NMF to be able to consider more of the local information of the data, which grants NMF an advantage over VCA, N-FINDR. The disadvantage of NMF is that it fails in guaranteeing that its results are the global minimum instead of the local minimum \cite{albright2006algorithms}.
\section{Experimental Protocol}
The purpose of this paper is to extend the unmixing methods, which were previously restricted in the field of hyperspectral imaging, to the field of multispectral imaging. 

\begin{figure}[hbt!]
  \centering
    \includegraphics[width=1\linewidth]{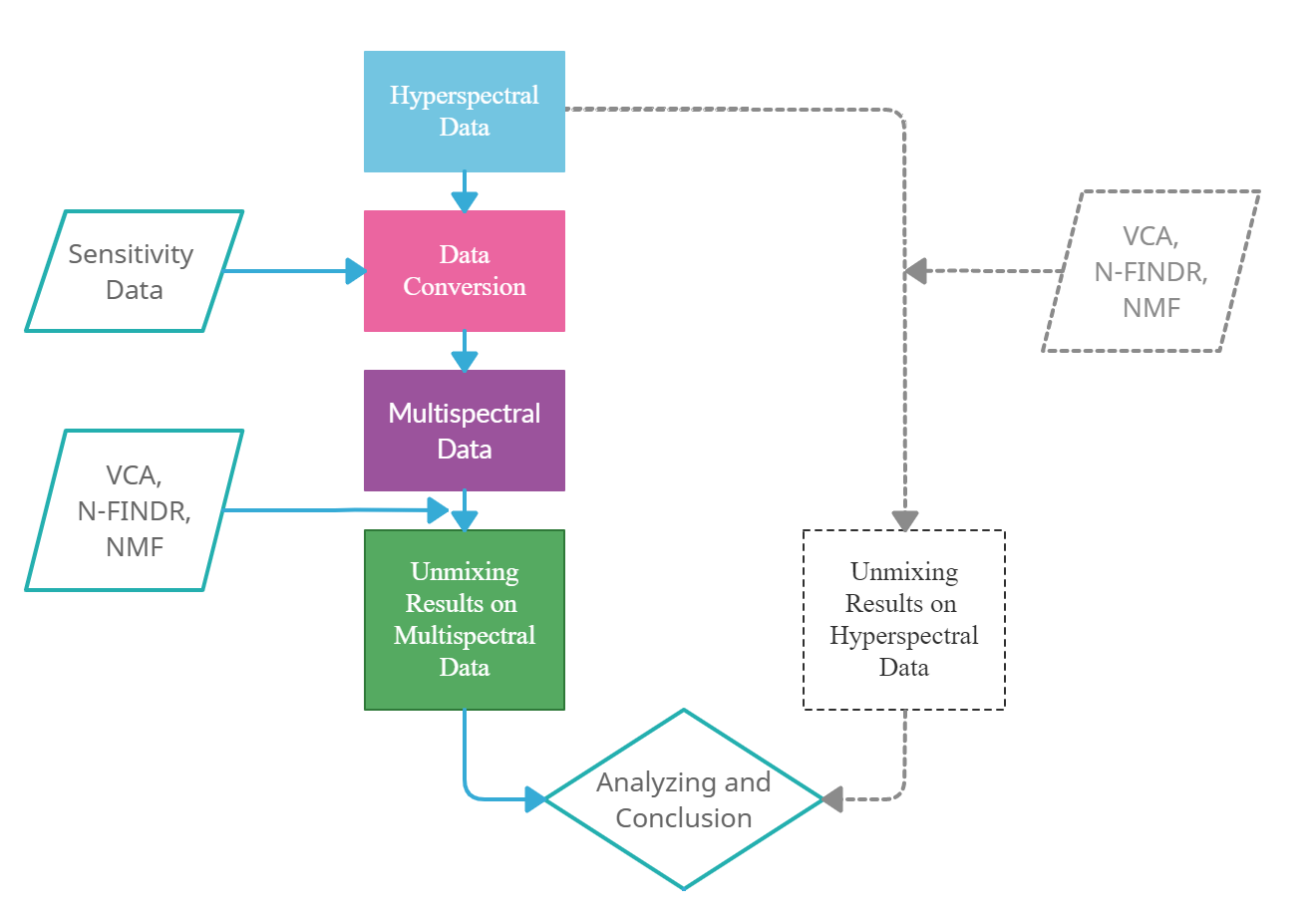}
  \caption{Experimental Procedure}
  \label{fig:experiment_procedure}
\end{figure}
\begin{figure}[hbt!]
  \centering
    \includegraphics[width=1\linewidth]{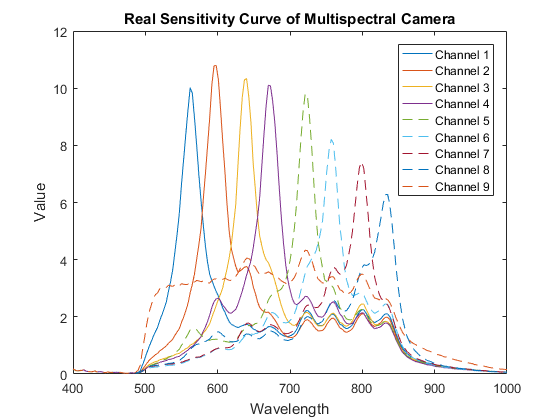}
  \caption{Real Sensitivity Curve of Multispectral Camera \\ (Channel 1-8: Selective channels. Channel 9: Panchromatic channel)}
  \label{fig:sensitivity_real}
\end{figure}
\begin{figure}[hbt!]
  \centering
    \includegraphics[width=1\linewidth]{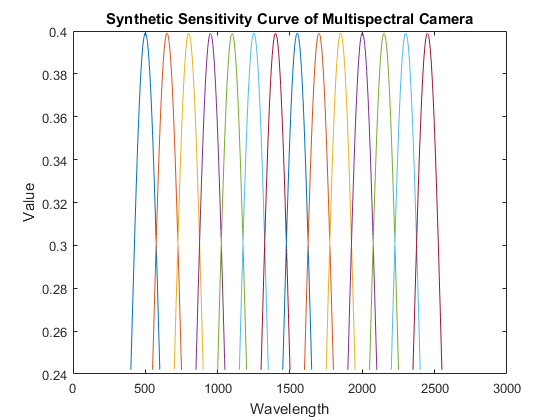}
  \caption{Synthetic Sensitivity Curve of Multispectral Camera}
  \label{fig:sensitivity_synthetic}
\end{figure}

As the Figure \ref{fig:experiment_procedure} suggests, we choose to simulate the multispectral images from hyperspectral images so that we can compare the results with the ground truth of hyperspectral imaging. In this paper, we have two different sensitivity curves. The first sensitivity curve is from a real multispectral camera (Figure \ref{fig:sensitivity_real}), while the second sensitivity curve is a synthetic curve (Figure \ref{fig:sensitivity_synthetic}). Their wavelength scopes are respectively 400--1000 nm, 400--2500 nm. Due to the page limit, we present the results of both sensitivity curves upon the first dataset, but only the result of real sensitivity curve upon the second dataset. 

From Equation \ref{eq:data_conversion}, we are able to obtain the multispectral imaging as if acquired from a multispectral camera. Equation \ref{eq:data_conversion} is computed for each channel, but for most cases, the $R(\lambda)$ and $S(\lambda)$ usually have different intervals. Therefore, the bandwidth of Sensitivity Data $S(\lambda)$ needs to be adapted to the spectrometer measure by linear interpolation.


\begin{equation} \label{eq:data_conversion}
  Y    =  \int_{\lambda_{min}}^{\lambda_{max}} I(\lambda)R(\lambda)S(\lambda) \,d\lambda ,
\end{equation}
where $I(\lambda)$is the Illumination, $R(\lambda)$ is the Reflectance of the Hyperspectral Data, $S(\lambda)$ is the Sensitivity Data of the Multispectral Camera, Y is the result of Multispectral Data, $\lambda_{min}$ is the minimum wavelength of the sensitivity curves, $\lambda_{max}$ is the maximum one. 


After obtaining the multispectral data, we apply the unmixing methods (VCA, N-FINDR, NMF) upon the multispectral data. Once the endmembers are extracted, we can calculate the abundance of each endmember.  


\section{Experiment and results}
In the following experimental part, we apply the methods to two datasets: Jasper dataset, Painting dataset. Our decision to choose these two datasets is based on two considerations. Firstly, Jasper is an open dataset in the field of remote sensing which has already been utilized and validated by researchers, like the cases in \cite{zhu2014effective,wang2015robust}. While at the same time, the Painting dataset is a dataset belonging to the field of pigments and painting. It is a new dataset that was published this year. Because these two datasets originate from two different domains, the experiment can better suggest and prove the possibilities for extending the unmixing methods for multispectral imaging. Secondly, we have ground truths for both of these two datasets, which will enable us to validate our results and methods.

\subsection{Experiment on Multispectral Data of Jasper}

The Jasper dataset is an open dataset taken with the AVIRIS camera. It is a hyperspectral image with 100 $\times$ 100 pixels and 198 bands from 400 nm to 2500 nm.  Four endmembers are concerned in this dataset,  they are respectively \textit{Tree}, \textit{Water}, \textit{Dirt} and \textit{Road}. The  ground-truth spectra and abundances of these four endmembers are already known. Using the radiance value, we can simulate the Multispectral image.

\begin{figure*}[hbt!]
  \centering
    \includegraphics[width=\textwidth]{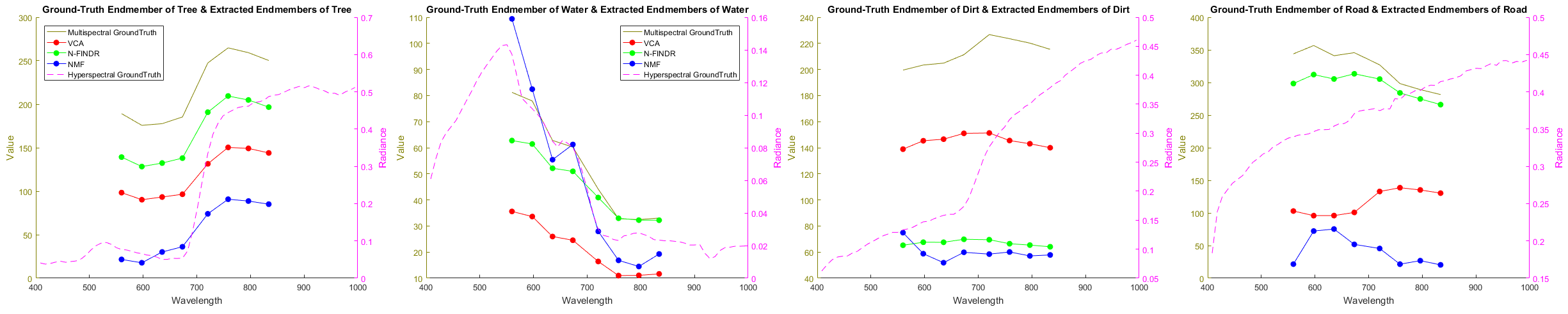}
  \caption{Unmixing results of Multispectral Jasper (using Real Sensitivity)}
  \label{fig:MultiJasper_ALL_comparison}
\end{figure*}

\begin{figure*}[hbt!]
  \centering
    \includegraphics[width=\textwidth]{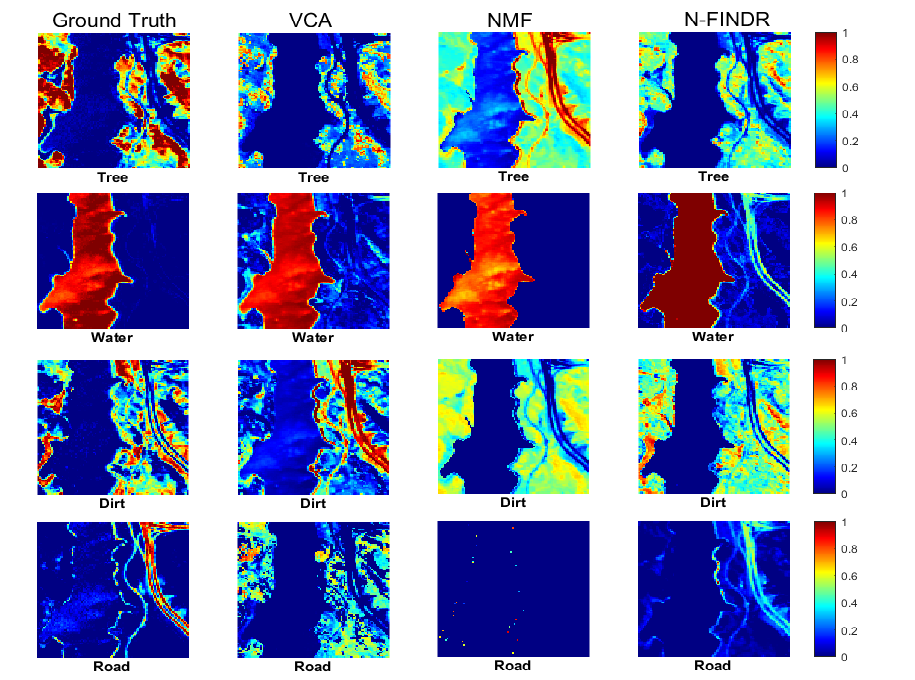}
  \caption{Abundance Maps of the Unmixing results of Multispectral Jasper (using Real Sensitivity)}
  \label{fig:MultiJasper_AbundanceMaps}
\end{figure*}

\begin{figure*}[hbt!]
  \centering
    \includegraphics[width=\textwidth]{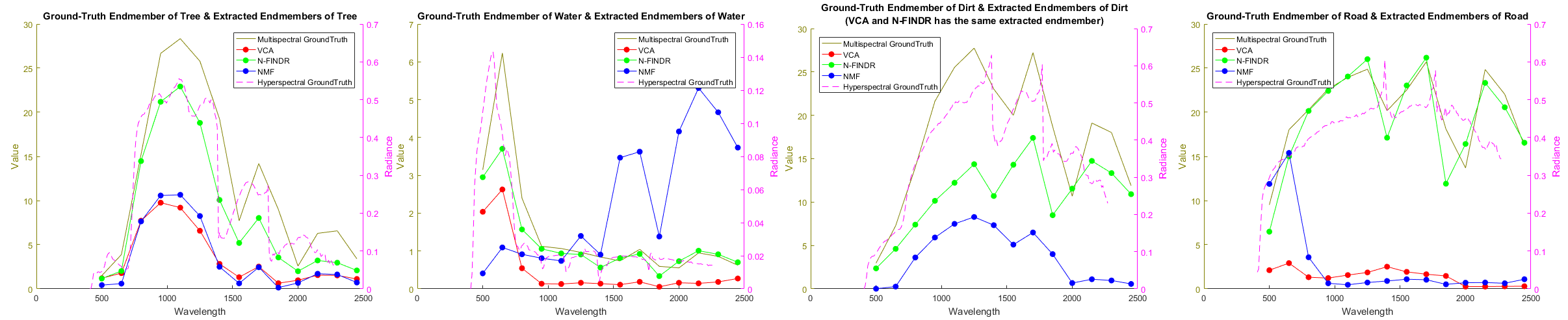}
  \caption{Unmixing results of Multispectral Jasper (using Synthetic Sensitivity)}
  \label{fig:SyntheticJasper_ALL_comparison}
\end{figure*}

\begin{figure*}[hbt!]
  \centering
    \includegraphics[width=\textwidth]{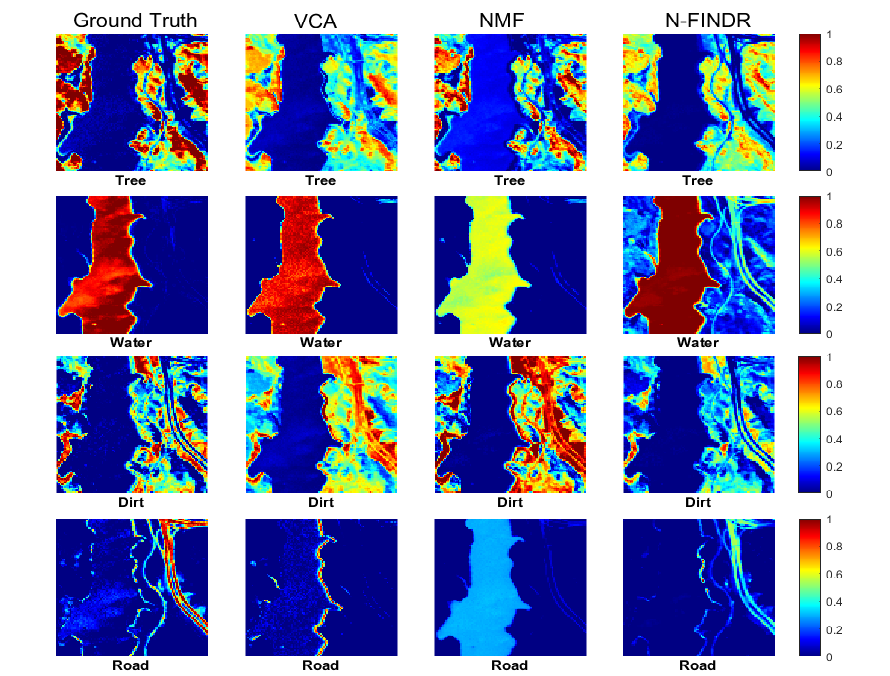}
  \caption{Unmixing results of Multispectral Jasper (using Synthetic Sensitivity)}
  \label{fig:SyntheticJasper_AbundanceMaps}
\end{figure*}

Following the Experimental Procedure showed in Figure \ref{fig:experiment_procedure}, we can have the unmixing results of Multispectral Jasper. In Figure \ref{fig:MultiJasper_ALL_comparison}, the Hyperspectral ground truth and Multispectral ground truth are demonstrated with the Unmixing results upon the Multispectral Jasper using VCA, N-FINDR, NMF. The Multispectral results are displayed with only 8 values. The first eight bands are relatively selective (Figure \ref{fig:sensitivity_real}). The ninth band of the Multispectral camera is associated to a panchromatic filter (Figure \ref{fig:sensitivity_real}) and so is not displayed on the wavelength spectra. To verify the consistency of the simulated image, we also displayed the Hyperspectral ground truth (It has the color of Magenta and corresponding vertical axis indicates the Radiance in the right side of the figure). The Multispectral ground truth has the color of yellow and corresponding vertical axis indicates the Value in the left side of the figure.

In Figure \ref{fig:MultiJasper_ALL_comparison}, we remark that the Multispectral Unmixing of VCA, N-FINDR, NMF and Multispectral ground truth have similar trends for these four endmembers. In fact, among these three unmixing methods, the results of N-FINDR are closer to the Multispectral ground truth for the \textit{Tree}, \textit{Water}, and \textit{Road} endmembers. At the same time, the results of NMF are the farthest from the Multispectral ground truth for the \textit{Tree}, \textit{Dirt}, and \textit{Road} endmembers. Summarized from the unmixing results of these four endmembers, we can find out that the N-FINDR obtains the best result in unmixing Multispectral Jasper, while the NMF performs the worst in unmixing Multispectral Jasper.

There exists another point in Figure \ref{fig:MultiJasper_ALL_comparison} that is worth noticing. When we compare the trends between Multispectral Jasper and Hyperspectral Jasper, their trends are very consistent for the \textit{Tree} and \textit{Water} endmembers. It seems that they are not congruent in the trend when it comes to the endmember of \textit{Road}. In fact, the trends of Hyperspectral ground truth and Multispectral ground truth are still consistent when we consider the fact that the peaks of sensitivity curves are lower at longer wavelengths (as Figure \ref{fig:sensitivity_real} shows). When applying Equation \ref{eq:data_conversion}, the trends of the multispectral results are decreasing, because they are more influenced by the decreasing peaks than the increase in radiance. This is the reason why in the \textit{Road} endmember, the Hyperspectral ground truth increases but the Multispectral ground truth decreases.

Figure \ref{fig:MultiJasper_AbundanceMaps} gives the corresponding abundance map of each endmember. When we examine the unmixing results of the endmember \textit{Road}, it is obvious that N-FINDR is the only method that is able to successfully detect the \textit{Road}. Both VCA and NMF have failed to obtain the \textit{Road} endmember. In fact, in the NMF unmixing, the \textit{Road} result does not contain any valuable information. Meanwhile, when we observe the unmixing results of \textit{Water}, it is evident that NMF performs the best, while N-FINDR mistakenly classifies parts of \textit{Road} as the \textit{Water}. Summarizing from Figure \ref{fig:MultiJasper_ALL_comparison} and Figure \ref{fig:MultiJasper_AbundanceMaps}, it is safe to say that N-FINDR performs the best, VCA the second, and NMF the worst when considering the Multispectral Jasper (using Real Sensitivity).

Because the Synthetic Sensitivity Curves have 14 bands, we can notice there are 14 values for each Multispectral results in Figure \ref{fig:SyntheticJasper_ALL_comparison}. N-FINDR has an excellent performance in detecting all the four endmembers. VCA performs worse, NMF performs the worst. This result is similar to the result in Figure \ref{fig:MultiJasper_ALL_comparison}, which indicates that the trends of unmixing results are consistent in spite of the differences of the Sensitivity Curves. Additionally, in Figure \ref{fig:SyntheticJasper_AbundanceMaps}, VCA and NMF have not be able to detect the endmember of \textit{Road} successfully, but N-FINDR performs well in finding \textit{Road}. Considering the fact that N-FINDR is only method detecting \textit{Road} correctly in Figure \ref{fig:MultiJasper_AbundanceMaps}, we can conclude that N-FINDR is the most robust among these three methods.

\subsection{Experiment on Multispectral Data of Painting}
This Painting dataset was firstly presented by Grillini \textit{et al.} \cite{grillini2021comparison}. In this dataset, seven different pigments are concerned: \textit{Kremer White}, \textit{Ultramarine Blue}, \textit{Naples Yellow}, \textit{Carmine}, \textit{Vermilion}, \textit{Viridian Green}, and \textit{Gold Ochre DD}. There are 175 painted patches in total. Each 2cm $\times$ 2cm patch is either a pure pigment or the known mixture of 2 to 3 pigments. According to the painting conservators, it is very rare to find in oil painting any mixture whose number of pigments exceeds 3. This dataset presents the reflectance of each patch from 400 nm to 1000 nm with 186 bands.

\begin{figure*}[hbt!]
  \centering
    \includegraphics[width=\textwidth]{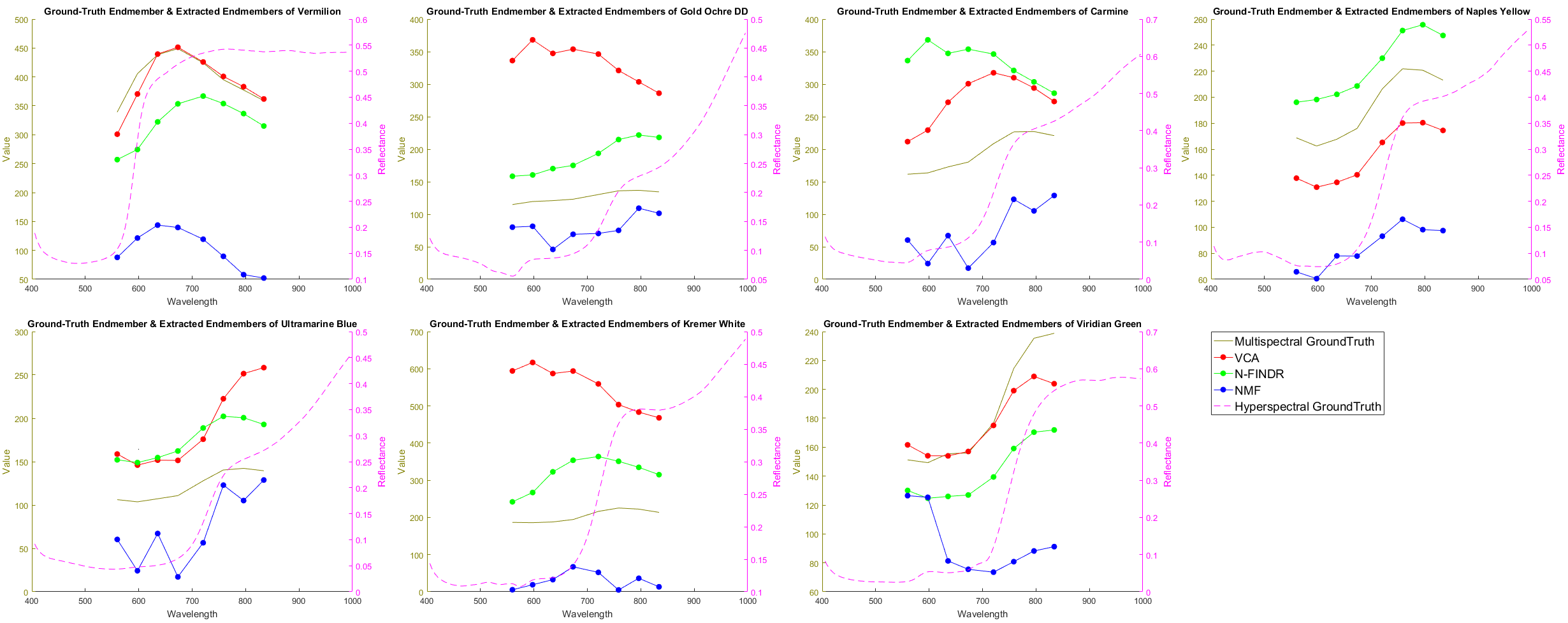}
  \caption{Unmixing results of Multispectral Painting}
  \label{fig:MultiPainting_ALL_comparison}
\end{figure*}

In Figure \ref{fig:MultiPainting_ALL_comparison}, the Multispectral ground truth, Hyperspectral ground truth, and the endmembers extracted with VCA, NMF, N-FINDR are displayed. For each pigment, its Multispectral ground truth and its endmember extraction results from VCA, N-FINDR, and NMF are presented. Among these three endmember extraction Methods, VCA performs the best in the endmember of \textit{Vermilion}, \textit{Carmine}, and \textit{Viridian Green}, while N-FINDR performs the best for the \textit{Kremer White} and \textit{Ultramarine Blue} endmembers. Their performances upon \textit{Naples Yellow} are similar. From Figure \ref{fig:MultiPainting_ALL_comparison}, we can notice that the trends of the unmixing results of VCA, NMF, and N-FINDR are mostly consistent with the trend of Multispectral ground truth. Like the case of Figure \ref{fig:MultiJasper_ALL_comparison}, there also exists a similar phenomenon that the trend of Hyperspectral Painting seems inconsistent with that of the Multispectral Painting. This is also due to the reason that the peaks and values of Sensitivity Curves become smaller with the increasing of wavelength.

Unlike the dataset of Multispectral Jasper, which is an image of 100 $\times$ 100 pixels, the data size of Painting dataset is relatively small. Thus, it is not possible to present the results of abundance estimation of the Painting dataset using Abundance Map, which is very intuitive. In view of that, we put forward a metric called SAVD (Sum of the Absolute Value of Difference between estimation abundance and ground truth abundance) for measuring its performance in Abundance Estimation. SAVD can be expressed as below:
\begin{equation}
    \centering
    \mathit{SAVD} = \sum\limits_{k=1}^{n}|\mathit{EA}_{k}- \mathit{GTA}_{k}|,
\end{equation}
In this formula, n signifies the the number of endmembers. $\mathit{EA}_{k}$ signifies the estimated abundance of $k^{th}$ endmember. $\mathit{GTA}_{k}$ signifies the ground truth abundance of $k^{th}$ endmember. SAVD is an index which can show the performance of Abundance Estimation directly. Its value scope is from 0 to 200\%, this is because SAVD denotes the sum of all the absolute values of its compositions. For example, the SAVD of a color patch would be 200\% if it only contained \textit{Carmine} yet was classified as \textit{Vermilion}.

\begin {table}[hbt!]
\caption {Table 1: The Unmixing results on Painting Dataset}  
\begin{center}
\begin{tabular} {|p{2.5cm}||p{1.3cm}|p{1.3cm}|p{1.5cm}|} 
 \hline
 \multicolumn{4}{|c|}{SAVD} \\
 \hline
 Endmember& VCA &NMF&N-FINDR\\
 \specialrule{.1em}{.05em}{.05em} 
 Vermilion          & 18.4\%    & 20.1\%    &   23.2\%\\
 \hline
 Gold Ochre DD      & 26.5\%    & 14.3\%    &   18.0\%\\
 \hline
 Ultramarine Blue   & 18.5\%    & 58.9\%    &   25.3\%\\
 \hline
 Kremer White       & 15.5\%    & 14.3\%    &   17.0\%\\
\hline
 Carmine            & 23.4\%    & 22.6\%    &   25.6\%\\
 \hline
 Naples Yellow      & 36.2\%    & 14.6\%    &   20.8\%\\
 \hline
 Viridian Green     & 13.3\%    & 14.0\%    &   26.7\%\\
\specialrule{.2em}{.1em}{.1em}  
\em{Average of 7 pigments}             & \textit{21.7\%}    & \textit{22.7\%}    &   \textit{22.4\%}\\
\specialrule{.2em}{.1em}{.1em} 
\end{tabular}

\label{tab:MultiPainting_SAVD} 
\end{center}
\end {table}

In Table \ref{tab:MultiPainting_SAVD}, each pigment exists in 65 color patches, thus the SAVD of each pigment is their corresponding average of SAVD in these 65 color patches. Firstly, the average of 7 pigments of SAVD of each Unmixing method is very close (VCA is 21.7\%, NMF is 22.7\%, N-FINDR is 22.4\%). This has revealed that the performances of VCA, NMF, N-FINDR are quite close. Secondly, it is observed that the values of N-FINDR among the seven pigments are more balanced (The standard deviations of VCA, NMF, and N-FINDR are respectively 7.80\%, 16.30\% and 3.86\%). Meanwhile, some extreme values exist in the unmixing results of VCA and NMF. For example, in VCA, the unmixing of \textit{Naples Yellow} is 36.2\%. In NMF, the unmixing of \textit{Ultramarine Blue} is 58.9\%. The endmember extracted by the NMF methods is very different from the ground truth which can explain why its abundance estimation is worse. At the same time, from Figure \ref{fig:MultiPainting_ALL_comparison}, even though the endmembers extracted using VCA are very close to the Multispectral Groundtruth, there are still extreme values of VCA in Table \ref{tab:MultiPainting_SAVD}. 

\begin{figure}[hbt!]
  \centering
    \includegraphics[width=1\linewidth]{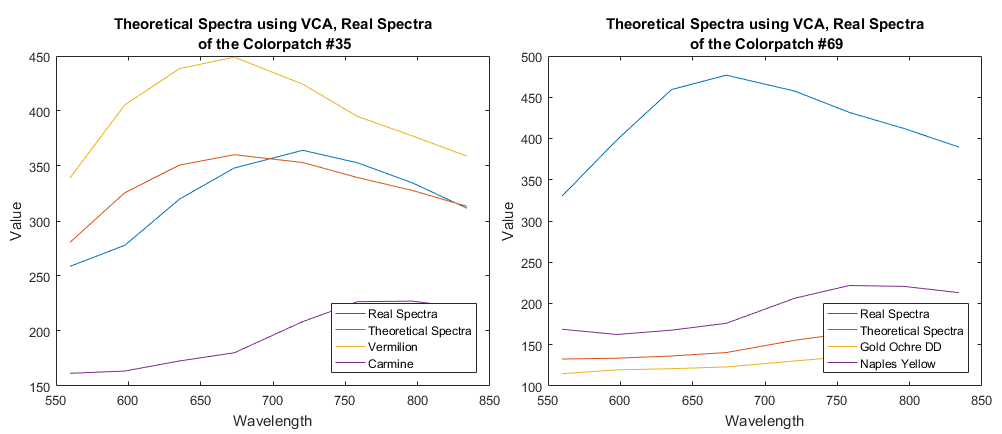}
  \caption{Theoretical Spectra using VCA and Real Spectra of the Colorpatches}
  \label{fig:MultiPainting_ALL_Colorpatch}
\end{figure}

To explore the possible reasons about these extreme values, we select two color patches: Colorpatch\#35, Colorpatch\#69. Such choice is based on the following criteria: Colorpatch\#35 has the smallest SAVD with VCA and Colorpatch\#69 has the highest. As VCA has the smallest SAVD, we choose to present only its results. In Figure \ref{fig:MultiPainting_ALL_Colorpatch}, the Real Spectra is the Multispectral value which is converted from the Reflectance acquired directly from spectrometer, while the Theoretical Spectra is calculated from the given spectra of the pigments and their corresponding abundances in each color patch. In the figure of Colorpatch\#69, we notice that the Real Spectra is very different from the Theoretical Spectra. This phenomenon is possibly caused by the utilization of linear Abundance Estimation. However, the mixing of oil in the painting is usually non-linear, which can account for the existence of extreme values of SAVD using VCA.


\section{Conclusion}
In this article, using two Sensitivity Curves of Multispectral Camera, we have been able to transform two Hyperspectral datasets (Jasper Dataset, Painting Dataset) into two Multispectral datasets. Then, we have applied three Unmixing methods (VCA, N-FINDR, NMF) upon these two Multispectral datasets. The results indicate that we can apply these Unmixing methods to multispectral datasets and obtain useful results. In the Jasper dataset, N-FINDR performs the best and NMF performs the worst. While in the Painting dataset, there does not exist very noticeable differences among these three methods. 

The results from our experiments have demonstrated the possibilities of these methods to be used in the application of unmixing the Multispectral images. Considering the fact that there exist other unmixing methods which are constrained to the field of Hyperspectral Unmixing (like the CNMF in \cite{zhang2015hyperspectral}), there exist the possibilities that they can also be extended to be applied to Multispectral Imaging in the future. 

This paper presents a proof of concept: unmixing methods can be applied to Multispectral data. Yet, we have only presented results from simulated Multispectral images. The next step will be to use real Multispectral images. Hyperspectral and Multispectral images present different characteristics (number and bandwidths of the bands). These differences should lead us to modify the unmixing method in order to adapt it to our specific data: Multispectral images. 


\section{Acknowledgement}
\begin{figure}[hbt!]
  \centering
    \includegraphics[width=1\linewidth]{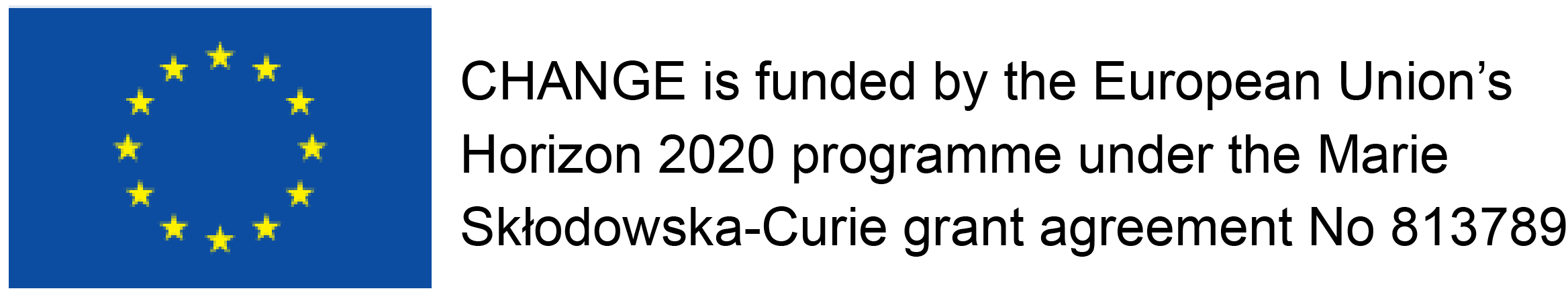}
\end{figure}




\end{document}